\documentclass[12pt,a4paper]{article}
\usepackage[english]{babel}
\usepackage{graphicx}

\newcommand{\alt}{\mathbin{\lower 3pt\hbox
   {$\rlap{\raise 5pt\hbox{$\char'074$}}\mathchar"7218$}}}
\newcommand{\agt}{\mathbin{\lower 3pt\hbox
   {$\rlap{\raise 5pt\hbox{$\char'076$}}\mathchar"7218$}}}

\begin{document}
\setcounter{footnote}{0}
\setcounter{equation}{0}
\setcounter{figure}{0}
\setcounter{table}{0}
\vspace*{5mm}

\begin{center}
{\large\bf On 't Hooft's representation of the $\beta$-function }

\vspace{4mm}
 I. M. Suslov \\
{ \it P.L.Kapitza Institute for Physical Problems,  119334 Moscow, Russia} \\
\vspace{6mm}

\begin{minipage}{135mm}
{\rm {\quad} It is demonstrated, that 't Hooft's renormalization scheme
(in which the $\beta$-function has exactly the two-loop form) is
generally  in conflict
with the natural physical requirements
and specifies the type of the field theory in an arbitrary
manner. It violates analytic properties in the coupling constant
plane and provokes misleading conclusion on accumulation of
singularities near the origin.
It artificially creates renormalon singularities, even if they
are absent in the physical scheme.  The 't Hooft scheme can be
used in the framework of perturbation theory but no global
conclusions should be drawn from it. } \end{minipage}
\end{center}

1. It is well-known, that the renormalization procedure is ambiguous
\cite{1,2}. Let for simplicity only the interaction constant $g$ is
renormalized.  Any observable quantity $A$, defined by a perturbation
expansion, is a function $F(g_0,\Lambda)$ of the bare value $g_0$ and the
momentum cut-off $\Lambda$.  According to the renormalization theory, $A$
becomes independent on $\Lambda$, if it is expressed in terms of
renormalized  $g$:
$$
A=F(g_0,\Lambda)=F_R(g)\,.
\eqno(1)
$$
The renormalized coupling constant $g$ is usually defined in
terms of a certain vertex, e.g. the four-leg vertex
$\Gamma_4(p_i,m)$ in the $g\phi^4$ theory, attributed to a
certain length scale $L$ through some choice of mass $m$ and
momenta $p_i$.  Two types of definition are conventionally used:

(1) $m$ is finite, $p_i=0$, and $g=\Gamma_4(0,m)$  corresponds to a length
scale  $L\sim m^{-1}$;

(2) $m=0$, $p_i\sim \mu$, and $g=\Gamma_4(p_i,0)$ corresponds to a length
scale $L=\mu^{-1}$; the condition $p_i\sim \mu$ is technically
realized by the equality
$$
p_i\cdot p_j =a_{ij} \mu^2 \,,
$$
where $a_{ij}$ are usually taken for the so called "symmetric point",
$a_{ij}=(4\delta_{ij} -1)/3$, though any other choice $a_{ij}\sim 1$
is possible.

Already the choice either (1) or (2) with different constants $a_{ij}$
provides essential  ambiguity of the renormalization scheme. In fact, the
physical condition that $g$ is determined by a vertex $\Gamma_4$ on the
length scale $L$ can be realized technically in many variants
(e.g. using averaging over $p_i$ with some weight function localized
on the scale $L^{-1}$)\footnote{\,The latter possibility is close
to a situation in the minimum subtraction (MS) scheme.
This scheme does not correspond to estimation of a certain vertex
for the specific choice of momenta. As  explained in the book
\cite{2a}, for any individual diagram one can choose
a scale $\lambda$ of order $\mu$, so that a usual subtraction on
the scale $\lambda$ is equivalent to the minimal subtraction
on the scale $\mu$. However, universal relation $\lambda=C\mu$
cannot be introduced because $C$ is different for different
diagrams. Nevertheless, $\lambda\sim\mu$ for any  diagram
simply on dimensional grounds. Consequently, the MS scheme
corresponds to a certain averaging over momenta on the scale
$\mu$. }.

On the conceptual level, the change of the renormalization scheme is simply a
change of variables,
$$
g=f(\tilde g) \,,
\eqno(2)
$$
transforming (1) into equation
$$
A=F(g_0,\Lambda)=\tilde F_R(\tilde g)
\eqno(3)
$$
of the same form. Such change of variables does not affect values
of observable quantities but changes a specific form of functions
$F_R(g)$.

In the lowest order of the perturbation expansion,
the equality $\Gamma_4=g_0$
takes place independently of $p_i$ and $m$, and one have  the first physical
restriction for the function $f(\tilde g)$:
\vspace{4mm}

$\qquad (R_1)\qquad\qquad\qquad\qquad\qquad f(\tilde g)=\tilde g + O(\tilde
g^2)\,.$
 \vspace{4mm}

\noindent In fact, the analogous condition $f(\tilde g)\sim \tilde
g$ should be valid in the order of magnitude in the large $\tilde
g$ region, so as to $\tilde g$ has the same physical sense, as $g$
(if for example $f(\tilde g)\sim \tilde g^2$, then $\tilde g$
corresponds to $\left(\Gamma_4 \right)^{1/2}$ instead $\Gamma_4$).
Consequently, the difference between the conventional
renormalization schemes corresponds to a change of variables (2)
with a function $f(\tilde g)$ of appearance shown in Fig.1.

\begin{figure}
\centerline{\includegraphics[width=4.8 in]{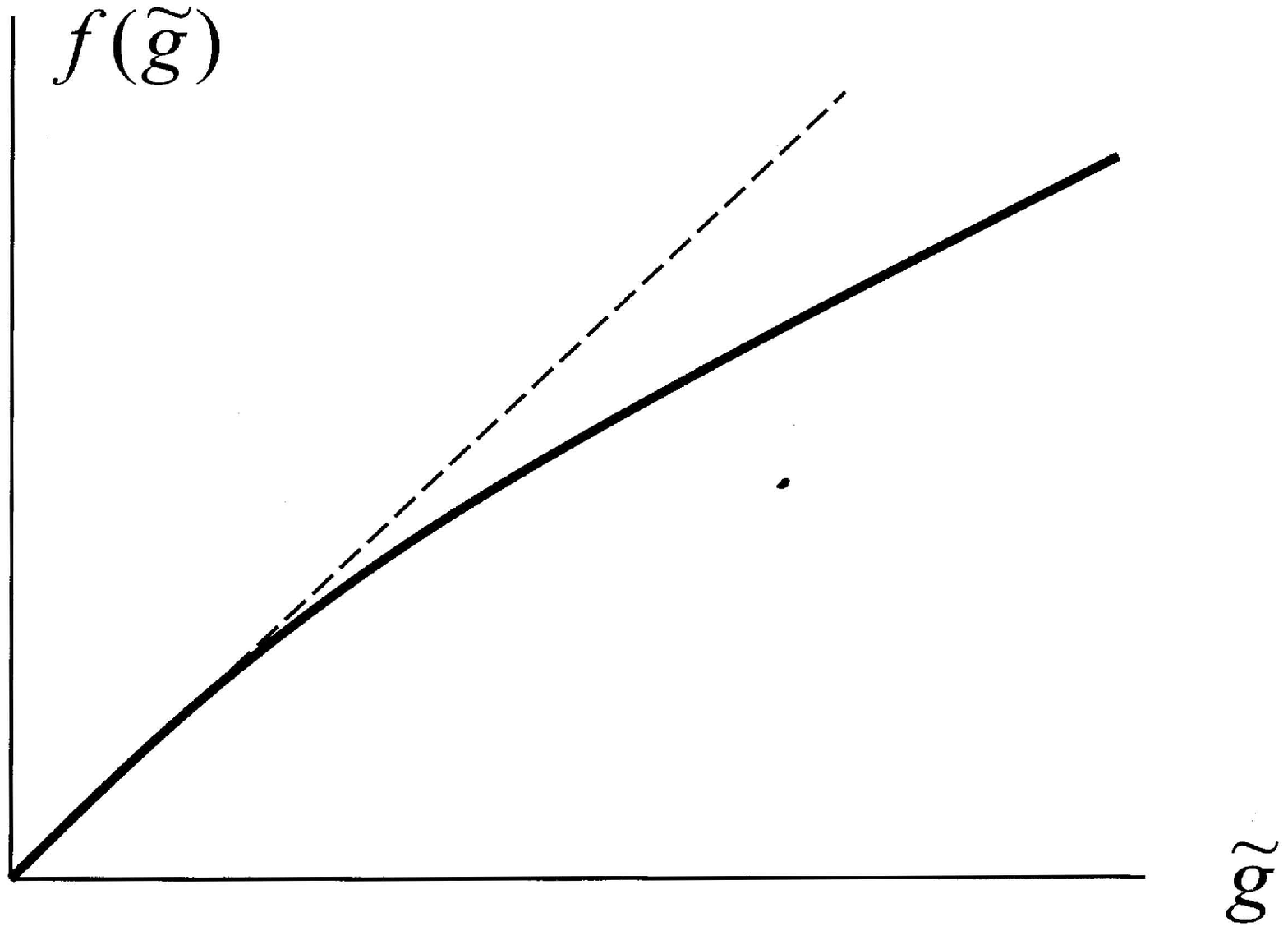}} \caption{}
\label{fig1}
\end{figure}

\vspace{3mm}

2. If we apply a change of variables (2) to the Gell-Mann -- Low
equation\,\footnote{\,Such form of equation corresponds literally to the
$\phi^4$ theory; in the case of QED and QCD one should use $e^2$ and $g^2$
instead of $g$ correspondingly.}
$$
-\frac{d g}{d\ln L} =\beta(g)=\beta_2 g^2 + \beta_3 g^3 +\beta_4
g^4 + \ldots \,, \eqno(4)
$$
then it transforms to
$$
-\frac{d \tilde g}{d\ln L} =
\tilde\beta(\tilde g),\qquad {\rm where}\qquad
\tilde\beta(\tilde g)=\beta(f(\tilde g))/f'(\tilde g)\,.
\eqno(5)
$$
It is easy to be convinced that the restriction $(R_1)$ provides invariance
of {\it two} coefficients $\beta_2$ and $\beta_3$ under the change of the
renormalization scheme.

In 1977 't Hooft has suggested \cite{3} to fix the renormalization scheme by
the condition, that Eq.5 has {\it exactly} the two-loop form
$$
-\frac{d \tilde g}{d\ln L} =\beta_2 \tilde g^2 + \beta_3 \tilde
g^3 \,.  \eqno(6)
$$
In the framework of  perturbation theory it
is always possible: if
$$
g=f(\tilde g)=\tilde g+\alpha_1 \tilde
g^2 +\alpha_2 \tilde g^3 +\alpha_3 \tilde g^4 +\ldots \,,
\eqno(7)
$$
then (5) has a form
$$
-\frac{d \tilde g}{d\ln L} =\beta_2 \tilde g^2 + \beta_3 \tilde
g^3 + (-\alpha_2 \beta_2+\beta_4) \tilde g^4 + (-2\alpha_3
\beta_2+\beta_5) \tilde g^5 + \ldots \eqno(8)
$$
The parameter
$\alpha_1$ can be fixed arbitrarily and we accepted $\alpha_1=0$
for simplicity. The coefficient $\alpha_n$ appears for the first
time in the term of the order $\tilde g^{n+2}$ and choosing successively
$$
\alpha_2=\frac{\beta_4}{\beta_2}\,,\qquad
\alpha_3=\frac{\beta_5}{2\beta_2}\,,\quad\ldots
\eqno(9)
$$
one can eliminate the terms $\tilde g^4$, $\tilde g^5\,\ldots$ in the
r.h.s. of (8). If this construction can be used
beyond perturbation context, it provides a powerful instrument
for investigation of general aspects of theory.

\vspace{3mm}

3. From the physical viewpoint, the choice of $f(\tilde g)$ is strongly
restricted (Fig.1), but formally one can choose this function rather arbitrary.
Nevertheless, there is a minimal physical restriction that should be added to
$(R_1)$:
\vspace{4mm}

$(R_2)\qquad$  $f(\tilde g)$  should be regular and provide  one to
one correspondence between $\phantom{vvvvvvvvvvv} g$ and $\tilde g$,
at least for their physical values, $g,\tilde g \in (0,\infty)$
\vspace{4mm}

\noindent
Indeed, variation of $g$ from $0$ to $\infty$ should correspond
to variation\,\footnote{\,The strong coupling region can be
physically inaccessible. In this case, the restriction $(R_2)$ can
be weakened:  variation of $g$ from $0$ to a finite value should
correspond to variation of $\tilde g$ from $0$ to a finite
value.}
of $\tilde g$ from $0$ to $\infty$, and this change of variables should not
create artificial singularities in the theory. It should be stressed, that
$(R_2)$ is not controlled in the above construction, where $f(\tilde g)$
is defined by a formal series in $\tilde g$. It is easy to
demonstrate, that restriction $(R_2)$ forbides to use
't Hooft's construction beyond perturbation theory.
\vspace{2mm}

According to classification by Bogolyubov and Shirkov \cite{1}, there are
three possible types of the $\beta$-function, corresponding to three
qualitatively different situations for the dependence of $g$ on
the length scale $L$. For $\beta_2>0$ they are:

\vspace{3mm}

(a) $\beta(g)$ has a nontrivial zero $g_c$; then $g(L)\to g_c$ for
$L\to 0$.

(b) $\beta(g)$ is nonalternating and behaves as $g^\alpha$ with $\alpha\le 1$
for $g\to\infty$; then $g(L)\to \infty$ for $L\to 0$.

(c) $\beta(g)$ is nonalternating and behaves as $g^\alpha$ with $\alpha> 1$
for $g\to\infty$; then $g(L)\to \infty$ at some finite point  $L_0$ (Landau
pole) and the dependence $g(L)$ is not defined for $L<L_0$, signalling that
the theory is internally inconsistent (or trivial).

\vspace{3mm}
\noindent
In the case $\beta_2<0$, the same conclusions hold for the limit $L\to\infty$
instead of $L\to 0$.
\vspace{2mm}

It is easy to see, that the restriction $(R_2)$ forbids to
transform one of the situations (a), (b), (c) into another. Let
$\tilde g$ corresponds to (a) or (b), and $g$ corresponds to (c):
then $g$ goes to infinity at the point $L=L_0$, where $\tilde g$
has a finite value $g^*$. Consequently, $f(\tilde g)\to\infty$
for $\tilde g\to g^*$ and regularity of $f(\tilde g)$ is
violated; more than that, $f(\tilde g)$ is not defined for
$\tilde g>g^*$ (Fig.2,a). Analogously, if $g$ corresponds to (b)
and $\tilde g$ corresponds to (a), then $g\to\infty$ and $\tilde
g\to g_c$ in the small $L$ limit; so $f(\tilde g)\to\infty$ for
$\tilde g\to g_c$ and $f(\tilde g)$ is not regular, while its
inverse is double-valued (Fig. 2,b). We see, that classification
by Bogolyubov and Shirkov has an absolute character and cannot be smashed
by the change of the renormalization scheme.
\begin{figure}
\centerline{\includegraphics[width=5.5 in]{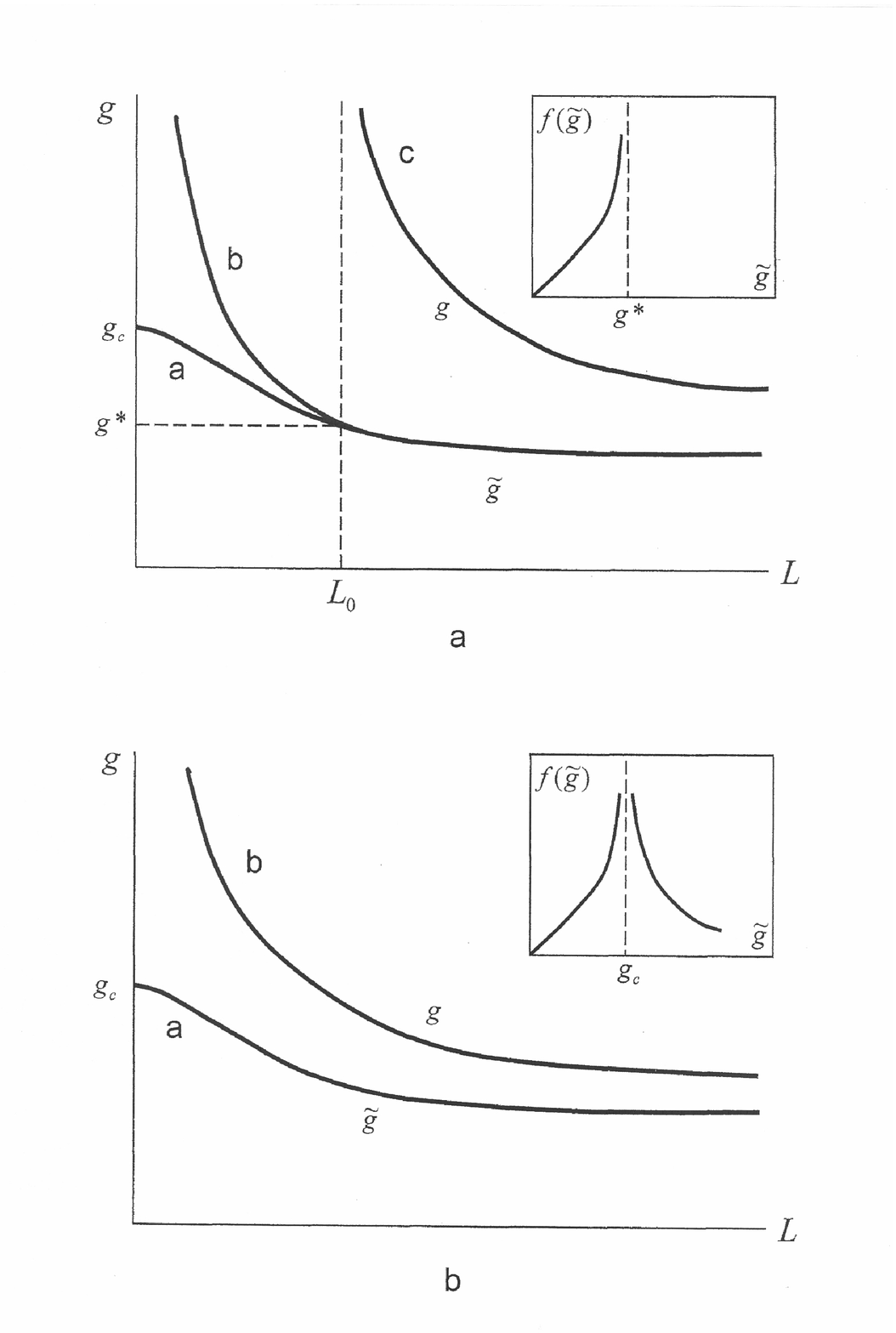}} \caption{}
\label{fig2}
\end{figure}

When 't Hooft's form (6) is postulated, a situation (b) becomes impossible
from the very beginning. The choice between other two situations is also
made, when the known coefficients $\beta_2$ and $\beta_3$ are taken into
account. Consequently, the type of the field theory is fixed, using the
knowledge of only two expansion coefficients, but that is
surely unjustifiable.  It easy to see, that 't Hooft's construction
predetermines internal inconsistency for QED ($\beta_2>0$, $\beta_3>0$) and QCD
($\beta_2<0$, $\beta_3<0$), and the fixed-point situation (a) for the $\phi^4$
theory ($\beta_2>0$, $\beta_3<0$).

It is commonly accepted that there no effective way beyond
perturbative theory. In fact, such way does exist. One can
calculate few first expansion coefficients diagrammatically
and their large-order asymptotics in the framework of the
instanton method suggested by Lipatov \cite{4}; producing the
smooth interpolation for the coefficient function,
one can find the sum of the whole
perturbation series. Such program was realized in \cite {5,6,7}
for reconstruction of the $\beta$-fuctions for the main field
theories (see also the review article \cite{8}). The results
have reasonable uncertainty and suggest a
situation (b) for the $\phi^4$ theory \cite{5} and QED \cite{6},
while situations (a) and (b) are possible for QCD \cite{7}.
All these results are in
conflict with 't Hooft's construction. Of course, one can have a
reasonable doubt that existing information is sufficient for reliable
reconstruction of the $\beta$-functions, but the results of \cite{5,6,7} are
certainly more reliable, than an arbitrary choice made in the 't
Hooft scheme.  In the case of the $\phi^4$ theory, there
is some controversy concerning the asymptotics of the
$\beta$-function \cite{5,9,10,11}, but there is a consensus that
the $\beta$-function is not alternating. The same conclusion
follows from the lattice results \cite{12} and the real-space renormalization
group analysis \cite{13}.\,\footnote{\,Usually these results are considered as
evidence of triviality of the $\phi^4$ theory, but in fact they demonstrate
only absense of the nontrivial zero for the $\beta$-function (see the detailed
discussion in \cite{5,8}). } As for QCD, it looks as successful theory of
strong interactions and hardly deserves a status of internally inconsistent
theory.

According to 't Hooft, an arbitrary $\beta$--function can be
reduced to the form (6). It  creates an illusion that
the physical $\beta$--function is not interesting quantity.
In fact, the latter has the fundamental significance,
allowing to distinguish three qualitatively different
types (a),(b),(c) of field theory. This question is not pure
academic.  For example, the conventional bound on the Higgs mass
is based on the expected triviality of the $\phi^4$ theory
\cite{20} and appears completely wrong, if it is not trivial.
The latter looks rather probable, according to \cite{5}.

\vspace{3mm}
4. In fact, singularity of $f(\tilde g)$ in the complex plane
is evident from the very beginning. It is clear from the Dyson
type arguments \cite{40} and instanton calculations \cite{4}
that perturbation series for $\beta(g)$ is factorially divergent
and $g=0$ is essential singularity; in fact, it is a branching
point and all quantities have at least two leafs of the
Riemann surface. In the 't Hooft scheme, $\beta$-function
is polynomial and does not possess the correct analytic
properties.

\vspace{3mm}

5. As immediate application of his scheme, 't Hooft
derived accumulation of singularities for the Green
functions near the origin $g=0$. He used the fact that
momentum $k$ enters all quantities in combination
$1/g+\beta_2 \ln(k^2/\mu^2)$.  On the physical grounds, Green
functions contain singularities for $g>0$, $k^2<0$, while for
$k^2>0$ one expect singularities at the points
$$
\frac{1}{g} = real + \beta_2 (2n+1)\pi i\, \qquad
n=0,\pm 1,\pm 2,\ldots.
\eqno(10)
$$
Existence of such singularities has fundamental significance,
since strong Borel summability of perturbative expansions
becomes impossible.

Attempt to generalize this conclusion to the arbirary
renormalization scheme was made by Khuri \cite{14}. His
analysis is based on expected regularity of the function
$g=f(\tilde g)$, relating 't Hooft's and some
other scheme, in a certain sector of the complex plane. However,
in proving this regularity Khuri discarded (as improbable)
the case when $\beta(g)$ has an infinite set of zeroes
accumulating near the origin. In fact, this case is not
improbable.  Consider the simplest (zero-dimensional)
version of the functional integral entering the $\phi^4$
theory
$$
F(g)=\int\limits_{-\infty}^{\infty}d\phi\,
{\rm e}^{-\phi^2-g\phi^4}=
{\textstyle\frac{1}{2}}\, g^{-1/2}\,{\rm e }^{1/8g}\,
K_{1/4}\left(\textstyle\frac{1}{8g}\right)
\eqno(11)
$$
Its relation with the Mac-Donald function $K_\nu(x)$
can be established by observation that $F(g)$ satisfies an
equation \cite{14a}
$$
4g^2F''+(8g+1)F'+{\textstyle\frac{3}{4}} F=0
\eqno(12)
$$
with the boundary condition $F(0)=\sqrt{\pi}$. It is easy to
show that the Mac-Donald function $K_\nu(z)$ has not zeroes
on the main leaf of the Riemann surface, but has zeroes
on the neighbouring leafs; for large $|z|$ they
are\,\footnote{\,It is easy to be convinced in validity of this
result, using the relation for the Airy function, $Ai(x)\sim
K_{1/3}\left({\textstyle\frac{2}{3}} x^{2/3}\right)$ or
$K_{1/3}\left({\textstyle\frac{2}{3}} t{\rm e}^{\pm 3\pi i/2}
\right) \sim Ai(-t^{2/3})$, and noticing that $Ai(x)$ has zeroes
for negative $x$.}
$$
z_s=-{\textstyle\frac{1}{2}}\ln(2\cos{\pi\nu}) + {\rm
e}^{\pm 3\pi i/2} \left( {\textstyle\frac{3\pi}{4}} +\pi
s\right)\,,\qquad s\,-\,integer
\eqno(13)
$$
One can see  from (11) that zeroes (13)
correspond to the points of kind (10) in the complex $g$ plane.
It is typical for functional integrals to have zeroes in
such points and it is not miraculous if $\beta(g)$ has also such
zeroes.

Then, according to Khuri's analysis, the function $g=f(\tilde g)$
is badly singular and has infinite number of singularities in
the points of type (10); hence, one cannot be sure, are
't Hooft's singularities (10) of physical relevance or they are
created by the singular transformation $g=f(\tilde g)$.

One can come to the problem from another side.
Zeroes of functional integrals correspond to
poles in the Green functions (which are determined by ratios
of such integrals), and hence their singularities are indeed of
type (10). However (!) they lie on unphysical leaf of the Riemann
surface. The choice of the leaf was not controlled in 't Hooft's
considerations, since his scheme does not reproduce the correct
analytic properties (Sec.4); in fact, such choice is not
trivial since the Stokes
phenomenon is intrinsic for functional integrals.

Regularity of the Green functions on the physical leaf can be
easily shown, if one accept that their Borel transforms have
the power-like behavior at infinity and suggest that
$\beta(g)\sim g^\alpha$, $\alpha\le 1$, for large $g$ \cite{16}.
Such assumptions look rather realistic according to
\cite{5,6,7}.

We see that conclusion on accumulation of singularities
following from the 't Hooft scheme appears to be misleading:
such singularities may either be absent or lie on  unphysical
leaf.

\vspace{3mm}

6. Another related aspect is the problem of renormalon
singularities in the Borel plane \cite{3,13a}. According to the
recent analysis \cite{16}, existence or absence of such
singularities is related with the analytic properties of the
$\beta$-function. Briefly, results are as follows:

(i) Renormalon singularities are absent, if $\beta(g)$ has a proper
behavior at infinity, $\beta(g)\sim g^\alpha$ with $\alpha\le 1$,
and its singularities at finite points $g_c$ are sufficiently weak,
so that $1/\beta(g)$ is not integrable at $g_c$
(i.e. $\beta(g)\sim (g-g_c)^\gamma$ with $\gamma\ge 1$).

(ii) Renormalon singularities exist, if at least one condition
named in (i) is violated.

It is easy to see, that 't Hooft's form (6) corresponds to
the behavior $\beta(g)\sim g^3$ at infinity and automatically
creates renormalon singularities, even if they were absent in
the physical renormalization scheme. It makes the field theory
to be ill-defined due to impossibility of the proper definition
of  functional integrals. Indeed, the classical definition of
the functional integral via the perturbation theory is
defective due to non-Borel-summability of the perturbative
series, while the lattice definition is doubtful due to
restriction of large momenta, which are responsible for
renormalon contributions \cite{8,16}.
Contrary, the results of \cite{5,6,7} show the possibility
of self-consistent elimination of renormalon singularities and formulation of
the well-defined field theory without renormalons \cite{8,16}.

{\qquad\qquad\qquad\qquad\qquad\qquad----------------------------------}

In conclusion, the 't Hooft representation for the
$\beta$-function (6) is generally in conflict with the natural
physical requirements and specifies the type of the field theory
in an arbitrary manner. It violates analytic properties in the
complex $g$ plane and provokes misleading conclusion on
accumulation of singularities near the origin.
  It artificially creates renormalon
 singularities, even if they are absent in the physical scheme.
The 't Hooft scheme can be used in the framework of perturbation
theory but no global conclusions should be drawn from it.

Author is indebted to participants of the PNPI Winter school for
stimulating discussions, and to F.V.Tkachov and A.L.Kataev for
consultations on the MS scheme.


\end{document}